\documentclass[aps,showpacs,preprintnumbers,nofootinbib]{revtex4}%
\usepackage{amssymb}
\usepackage{amsmath}
\usepackage{bm}
\usepackage{amsfonts}
\usepackage{graphicx}%
\providecommand{\U}[1]{\protect\rule{.1in}{.1in}}
\begin{document}
\title{(2+1)-Dimensional Gravity in Weyl Integrable Spacetime}
\author{$^{1}$ J. E. Madriz Aguilar\thanks{E-mail address: jemadriz@fisica.ufpb.br},
$^{2}$ C. Romero\thanks{E-mail address: cromero@pq.cnpq.br} \ \ $^{2}$J.B.
Fonseca Neto\thanks{jfonseca@fisica.ufpb.br} , $^{2}$T. S. Almeida and J. B.
Formiga$^{3}$}
\affiliation{$^{1}$ Instituto de F\'isica de la Universidad de Guanajuato, C.P. 37150,
Le\'on Guanajuato, M\'exico,}
\affiliation{and}
\affiliation{Departamento de F\'isica, Facultad de Ciencias Exactas y Naturales,
Universidad Nacional de Mar del Plata, Funes 3350, C.P. 7600, Mar del Plata, Argentina.}
\affiliation{}
\affiliation{$^{2}$ Departamento de F\'{\i}sica, Universidade Federal da Para\'{\i}ba,
Caixa Postal 5008, 58059-970 Jo\~{a}o Pessoa, PB, Brazil}
\affiliation{\ $^{3}$ Centro de Ci\^{e}ncias da Natureza, Universidade Estadual do
Piau\'{\i}, Caixa Postal 381, 64002-150 Teresina. PI, Brazil }
\affiliation{E-mail: cromero@fisica.ufpb.br}

\begin{abstract}
We investigate (2+1)-dimensional gravity in a Weyl integrable spacetime
(WIST). We show that, unlike general relativity, this scalar-tensor theory has
a Newtonian limit \ for any dimension $n\geqslant3$ and that in three
dimensions the congruence of world lines of particles of a pressureless fluid
has a non-vanishing geodesic deviation. We present and discuss a\ class of
static vacuum solutions generated by a circularly symmetric matter
distribution that for certain values of the parameter $\omega$ corresponds to
a space-time with a naked singularity at the center of the matter
distribution. We interpret all these results as being a direct consequence of
the space-time geometry.

\end{abstract}

\maketitle

\vskip .5cm


\section{Introduction}

During the past three decades a great deal of effort has gone to the
investigation into gravity theories in (2+1) space-time dimensions \cite{libro
LDG}. It seems that part of this interest has been motivated by the fact that
in this dimensionality Einstein gravity presents some odd peculiarities.
First, it has no propagating gravity modes, which implies that its quantum
version contains no gravitons. This is due to the fact that in (2+1)
dimensions space-time is flat outside sources. For point sources gravity
manifests itself as global topological defects rather than local geometrical
curvature. Secondly, in three-dimensional space-time Einstein theory does not
reduce to Newtonian gravity in the static weak-field regime \cite{Giddings}.
Another interest\ in (2+1)- dimensional gravity comes from the fact that
studying classical models in lower dimensions has often been helpful to
understand their quantum version \cite{carlip}.

The failure of (2+1)-dimensional Einstein gravity to provide a relativistic
generalization of \ two-dimensional Newtonian gravity has led some authors to
investigate the same problem in other theories of gravity. It has been proved
that at least in two distinct gravitational theories the Newtonian limit may
be recovered. These are the Brans-Dicke theory and the teleparallel gravity, a
metric theory in which gravity is purely due to torsion \cite{BD}. Another
approach to address this question is to consider (2+1)- dimensional gravity as
obtained through a dimensional reduction of four-dimensional Einstein gravity
\cite{Verbin}. The basic motivation in many of these attempts is to seek
alternative ways beyond the scope of general relativity to gain more insight
into some problems that do not seem to have a satisfactory solution in the
context of Einstein theory. With the same idea in mind we consider the subject
in the context of another theory of gravity, namely, the Weyl integrable
space-time theory (WIST) \cite{Novello}. In this approach, the geometry of
space-time is not Riemannian, \ but corresponds to what is called a Weyl
integrable geometry, a particular version of the geometry developed by H. Weyl
in 1918 in connection with his gravitational theory \cite{Weyl}. We show that,
in addition to leading to a Newtonian limit, WIST in (2+1) dimensions presents
some interesting properties that are not shared by Einstein theory, such as
geodesic deviation between particles in a dust distribution and\textit{\ }the
existence of naked singularities\textit{. }

The gravitational theory developed by Weyl arose as one of the first attempts
to unify gravity and electromagnetism. Its geometrical structure is based on
one of the simplest generalizations of Riemannian geometry. As is well known,
the kind of geometrical structure conceived by H. Weyl, although admirably
elegant and ingenious, turned out to be unsuitable as a physical theory. It
came from Einstein the first objections to the theory, who argued that in a
non-integrable\ Weyl geometry the existence of sharp spectral lines in the
presence of an electromagnetic field would not be possible since atomic clocks
would depend on their past history \cite{Pauli}. Later, it was found that a
variant of Weyl geometries, known as Weyl integrable geometry, does not suffer
from the drawback pointed out by Einstein, and since then has attracted the
attention of some cosmologists \cite{Novello}. In the opinion of some authors,
Weyl theory \ "contains a suggestive formalism and may still have the germs of
a future fruitful theory " \cite{Bazin}.

The paper is organized as follows. In Sections 2 and 3 we give a brief account
of the Weyl geometry and introduce the formalism of a specific theory of
gravity known \ as Weyl integrable space-time (WIST). The (2+1)-dimensional
version of this theory is presented in Section 4. In Section 5, we show that
WIST has a Newtonian limit for any dimension $n\geqslant3$. We proceed to
Section 6 to prove that, unlike (2+1)-dimensional general relativity, in
three-dimensional Weyl gravity the congruence of world lines of particles of a
pressureless fluid has a non-vanishing geodesic deviation. In Section 7, we
present and discuss a\ class of static vacuum solutions generated by a
circularly symmetric matter distribution. Finally, we conclude with some
remarks in Section 8.

\section{\bigskip Weyl geometry}

We can easily define Weyl geometry by its essential difference from the
Riemann geometry: while in the latter one makes the assumption that the
covariant derivative of the metric tensor $g$ is zero, in Weyl geometry we have%

\begin{equation}
\nabla_{\alpha}g_{\beta\lambda}=\sigma_{\alpha}g_{\beta\lambda}
\label{compatibility}%
\end{equation}
where $\sigma_{\alpha}$ \ denotes the components of a one-form field
$\sigma\ $with respect to a local coordinate basis \footnote{Throughout this
paper our convention is that Greek indices take values from $0$ to $n-1,$ $n$
being the space-time dimension.}. This represents a generalization of the
Riemannian condition of compatibility between the connection $\nabla$ and $g,$
which is equivalent to require the length of a vector to remain unaltered by
parallel transport \cite{Pauli}. If $\sigma$ $=d\phi,$ where $\phi$ is a
scalar field, then we have what is called\ an integrable Weyl geometry. A
differentiable manifold $M$ endowed with a metric $g$ and a Weyl scalar field
$\phi$ $\ $is usually referred to as a \textit{Weyl frame}. It is interesting
to note that the Weyl condition (\ref{compatibility}) remains unchanged\ when
we go to another Weyl frame $(M,\overline{g},\overline{\phi})$ by performing
the following simultaneous transformations in $g$ and $\phi$:%
\begin{equation}
\overline{g}=e^{f}g, \label{conformal}%
\end{equation}%
\begin{equation}
\overline{\phi}=\phi+f, \label{gauge}%
\end{equation}
where $f$ is a scalar function defined on $M$.

Quite analogously to Riemann geometry, the condition (\ref{compatibility}) is
sufficient to determine\ the Weyl connection\ $\nabla$\ in terms of the metric
$g$\ and the Weyl one-form field $\sigma.$ Indeed, a straightforward
calculation shows that one can express the components of the affine connection
with respect to an arbitrary vector basis completely in terms of the
components of $g$ and $\sigma$:%
\begin{equation}
\Gamma_{\beta\lambda}^{\alpha}=\{_{\beta\lambda}^{\alpha}\}-\frac{1}%
{2}g^{\alpha\mu}[g_{\mu\beta}\sigma_{\lambda}+g_{\mu\lambda}\sigma_{\beta
}-g_{\beta\lambda}\sigma_{\mu}], \label{Weylconnection}%
\end{equation}
where $\{_{\beta\lambda}^{\alpha}\}$ represents the Christoffel symbols.

A clear geometrical insight into the properties of Weyl parallel transport is
given by the following proposition: Let $M$ be a differentiable manifold with
an affine connection $\nabla$, a metric $g$ and a Weyl field of one-forms
$\sigma$. If $\nabla$ is compatible with $g$ in the Weyl sense,\ i.e. if
(\ref{compatibility}) holds, then for any smooth curve $\alpha=\alpha
(\lambda)$ and any pair of two parallel transported vector fields $V $ and $U$
along $\alpha,$ we have
\begin{equation}
\frac{d}{d\lambda}g(V,U)=\sigma(\frac{d}{d\lambda})g(V,U)
\label{covariantderivative}%
\end{equation}
where $\frac{d}{d\lambda}$ denotes the vector tangent to $\alpha$.

If we integrate the above equation along the curve $\alpha$, starting from a
point $P_{0}=\alpha(\lambda_{0}),$ then we obtain%
\begin{equation}
g(V(\lambda),U(\lambda))=g(V(\lambda_{0}),U(\lambda_{0}))e^{\int_{\lambda_{0}%
}^{\lambda}\sigma(\frac{d}{d\rho})d\rho} \label{integral}%
\end{equation}
Putting $U=V$ and denoting by $L(\lambda)$ the length of the vector
$V(\lambda)$ at an arbitrary point\ $P=\alpha(\lambda)$\ of the curve, then it
is easy to see that in a local coordinate system $\left\{  x^{\alpha}\right\}
$ the equation (\ref{covariantderivative}) reduces to
\[
\frac{dL}{d\lambda}=\frac{\sigma_{\alpha}}{2}\frac{dx^{\alpha}}{d\lambda}L
\]

Consider the set of all closed curves $\alpha:[a,b]\in R\rightarrow M$, i.e,
with $\alpha(a)=\alpha(b).$ Then, \ we have the equation
\[
g(V(b),U(b))=g(V(a),U(a))e^{\int_{a}^{b}\sigma(\frac{d}{d\lambda})d\lambda}.
\]
Now, it is the integral\ $^{\int_{a}^{b}\sigma(\frac{d}{d\lambda})d\lambda}$
that is responsible for the difference between the readings of two identical
atomic clocks following different paths. It follows from Stokes' theorem that
if $\sigma$ is an exact form, that is, \ if there exists a scalar function
$\phi$, such that $\sigma=d\phi$, then%

\[
\oint\sigma(\frac{d}{d\lambda})d\lambda=0
\]
for any loop. In other words, in this case the integral $e^{\int_{\lambda_{0}%
}^{\lambda}\sigma(\frac{d}{d\rho})d\rho}$ does not depend on the path.\ Since
it is this integral that regulates the way atomic clocks run this variant of
Weyl geometry does not suffer from the flaw pointed out by Einstein, and we
have what is often called in the literature a \textit{Weyl integrable
manifold}. In the next section we shall consider a theory of gravity
(WIST)\ formulated in a Weyl integrable manifold.

\section{Weyl integrable space-time theory in n dimensions}

In Weyl integrable\ space-time theory it is assumed\ that, in $n$ dimensions,
the dynamics of the gravitational field\ is given by the following action
\cite{Novello}:
\begin{equation}
^{(n)}\mathcal{S}=\int d^{n}x\,\sqrt{\left\vert g\right\vert }\left[
\!\mathcal{R}+\omega\phi_{,\alpha}\phi^{,\alpha}+\kappa_{n}e^{-n\phi/2}%
L_{m}\right]  \label{weyldynamics}%
\end{equation}
where $\omega$ is an arbitrary coupling constant, $\phi_{,\alpha}$ denotes the
derivative $\frac{\partial\phi}{\partial x^{\alpha}}$ of the Weyl scalar field
$\phi$, $\!\mathcal{R}$ is the Weylian Ricci scalar, $L_{m}$ is the Lagrangian
of matter, $\kappa_{n}$ denotes the Einstein constant in $n$ dimensions and,
as usual, $|g|$ indicates the absolute value of the determinant of the metric
tensor\ \footnote{Throughout this paper we shall adopt the following
convention in the definition of the Riemann and Ricci tensors: $R_{\;\mu
\beta\nu}^{\alpha}=\Gamma_{\beta\mu,\nu}^{\alpha}-\Gamma_{\mu\nu,\beta
}^{\alpha}+\Gamma_{\rho\nu}^{\alpha}\Gamma_{\beta\mu}^{\rho}-\Gamma_{\rho
\beta}^{\alpha}\Gamma_{\nu\mu}^{\rho};$ $R_{\mu\nu}=R_{\;\mu\alpha\nu}%
^{\alpha}.$ In this convention, we shall write the Einstein equations as
$R_{\mu\nu}-\frac{1}{2}Rg_{\mu\nu}=-\kappa T_{\mu\nu},$ with $\kappa
=\frac{8\pi G}{c4}$.}. It is important to note here that $L_{m}$ is
constructed by following the so-called Weyl minimal coupling prescription
\cite{Romero}. This means that it will be assumed that $L_{m}$ depends on
$\phi,$ $g_{\mu\nu}$ and the matter fields, here \ generically designated\ by
$\xi$, its form being obtained from the special theory of relativity through
the "minimum coupling" prescription $\eta_{\mu\nu}\rightarrow e^{-\phi}%
g_{\mu\nu}$ and $\partial_{\mu}\rightarrow\nabla_{\mu}$, where $\nabla_{\mu}$
denotes the covariant derivative with respect to the Weyl affine connection.
If we designate the Lagrangian of the matter fields in special relativity by
$L_{m}^{sr}=$ $L_{m}^{sr}(\eta,\xi,\partial\xi$ $)$, then the form of $L_{m}$
will be given by the rule $L_{m}(g,\phi,\xi,\nabla\xi)\equiv L_{m}%
^{sr}(e^{-\phi}g,\xi,\nabla\xi)$.\ As it can be easily seen, these rules also
ensure the invariance under Weyl transformations of part of the action that is
responsible for the coupling of matter with the gravitational field, and, at
the same time, reproduce the principle of minimal coupling adopted in general
relativity when we set $\phi=0$, that is, when we go to the Riemann frame by a
Weyl transformation.

We can also express the above action as%

\[
^{(n)}\mathcal{S}=\int d^{n}x\sqrt{|g|}\left[  R+\frac{(n-1)(n-2)+4\omega}%
{4}\phi_{,\alpha}\phi^{,\alpha}+\kappa_{n}e^{-n\phi/2}L_{m}\right]  ,
\]
where$\ R$ represents the scalar curvature evaluated with respect to the
Riemannian connection \footnote{In order to keep the kinetic term in the
Lagrangian we shall assume throughout the paper that $\omega\neq
-\frac{(n-1)(n-2)}{4}.$}. Let us now recall how the energy-momentum tensor
$T_{\mu\nu}(\phi,g,\xi$ $,\nabla\xi$ $)$ is defined in WIST gravity. In an
arbitrary Weyl frame $T_{\mu\nu}(\phi,g,\xi$ $,\nabla\xi$ $)$ is defined by
the formula
\begin{equation}
\delta\int d^{n}x\sqrt{|g|}e^{-n\phi/2}L_{m}(g_{\mu\nu},\phi,\xi,\nabla
\xi)=\int d^{n}x\sqrt{|g|}e^{-n\phi/2}T_{\mu\nu}(\phi,g_{\mu\nu},\xi,\nabla
\xi)\delta(e^{\phi}g^{\mu\nu}), \label{energy 1}%
\end{equation}
where the variation on the left-hand side must be carried out simultaneously
with respect to both $g_{\mu\nu}$ and $\phi.$ In order to see that the above
definition makes sense, it must be understood that the left-hand side of the
equation (\ref{energy 1}) can always be put in the same form of the right-hand
side of the same equation. This can easily be seen from the fact that $\delta
L_{m}=\frac{\partial L_{m}}{\partial g^{\mu\nu}}\delta g^{\mu\nu}%
+\frac{\partial L_{m}}{\partial\phi}\delta\phi=\frac{\partial L_{m}}%
{\partial(e^{\phi}g^{\mu\nu})}\delta(e^{\phi}g^{\mu\nu})$ and that
$\delta(\sqrt{|g|}e^{-n\phi/2})=-\frac{1}{2}\sqrt{|g|}e^{-\phi(1+n/2)}%
g_{\mu\nu}\delta(e^{\phi}g^{\mu\nu}).$

Varying the action $^{(n)}\mathcal{S}$\ with respect to the metric
$g_{\alpha\beta}$ and to the Weyl scalar field $\phi$, we obtain,
respectively, the following equations:
\begin{equation}
R_{\mu\nu}-\frac{1}{2}\,g_{\mu\nu}R+\frac{(n-1)(n-2)+4\omega}{4}\left[
\phi_{,\mu}\phi_{,\nu}-\frac{1}{2}g_{\mu\nu}\phi_{,\alpha}\phi^{,\alpha
}\right]  =-\kappa_{n}T_{\mu\nu}e^{(1-n/2)\phi}, \label{Weyl-n}%
\end{equation}

\begin{equation}
\square\phi=\frac{2\kappa_{n}}{(n-1)(n-2)+4\omega}e^{(1-n/2)\phi}T
\label{Weyl-n2}%
\end{equation}
where the symbol $^{(n)}\square$ denotes the $n$-dimensional D'Alembertian
operator with respect to the Riemannian connection, and $T=g^{\mu\nu}T_{\mu
\nu}$.

\section{Weyl integrable theory of gravity in (2+1)- dimensional space-time}

Let us now consider WIST in a (2+1)-dimensional space-time in the absence of
matter. In this case, if $n=3$ (\ref{Weyl-n})\ and (\ref{Weyl-n2}) become%

\begin{equation}
R_{\mu\nu}-\frac{1}{2}\,g_{\mu\nu}R+\frac{(1+2\omega)}{2}\left[  \phi_{,\mu
}\phi_{,\nu}-\frac{1}{2}g_{\mu\nu}\phi_{,\alpha}\phi^{,\alpha}\right]  =0,
\label{einstein-equations}%
\end{equation}%
\begin{equation}
\square\phi=0\text{ ,} \label{fi}%
\end{equation}
recalling that $\omega$ $\neq$ $-\frac{1}{2}.$On the other hand, if we take
the trace of the equation (\ref{Weyl-n})\ we\ shall get
\begin{equation}
R=-\frac{1}{2}\,(2\omega+1)\phi_{,\alpha}\phi^{,\alpha}. \label{a5}%
\end{equation}
Substituting (\ref{a5}) into equation (\ref{einstein-equations}) leads to
\begin{equation}
R_{\mu\nu}=-\frac{1}{2}\,(2\omega+1)\phi_{,\mu}\phi_{,\nu}. \label{a6}%
\end{equation}
Let us now recall the well-known mathematical fact which says that for $n>3$
the Riemann tensor $R_{\lambda\mu\nu\kappa}$\ can be decomposed in terms of
the Ricci tensor $R_{\mu\nu}$ the scalar curvature $R$\ and the Weyl tensor
$W_{\lambda\mu\nu\kappa}$. \ However, if $n=3$, then, because $W_{\lambda
\mu\nu\kappa}$ \ vanishes identically, we have the following expression for
$R_{\lambda\mu\nu\kappa}$ \cite{Weinberg}:
\begin{equation}
R_{\lambda\mu\nu\kappa}=g_{\lambda\nu}R_{\mu\kappa}-g_{\lambda\kappa}R_{\mu
\nu}+g_{\mu\kappa}R_{\lambda\nu}-g_{\mu\nu}R_{\lambda\kappa}-\frac{R}%
{2}\left(  g_{\lambda\nu}g_{\mu\kappa}-g_{\lambda\kappa}g_{\mu\nu}\right)  .
\label{Riemann}%
\end{equation}
\ By taking into account (\ref{a5}) and (\ref{a6}) we can express
(\ref{Riemann}) as%

\begin{equation}
R_{\lambda\mu\nu\kappa}=k_{n}e^{-\phi/2}\left[  g_{\lambda\kappa}T_{\mu\nu
}+g_{\mu\nu}T_{\lambda\kappa}-g_{\lambda\nu}T_{\mu\kappa}-g_{\mu\kappa
}T_{\lambda\nu}+\left(  g_{\lambda\nu}g_{\mu\kappa}-g_{\lambda\kappa}g_{\mu
\nu}\right)  T\right]  +\Psi_{\lambda\mu\nu\kappa} \label{a8}%
\end{equation}
where%

\begin{equation}
\Psi_{\lambda\mu\nu\kappa}=-\frac{1}{2}\,(2\omega+1)\left[  g_{\lambda\nu}%
\phi_{,\mu}\phi_{,\kappa}+g_{\mu\kappa}\phi_{,\lambda}\phi_{,\nu}%
-g_{\lambda\kappa}\phi_{,\mu}\phi_{,\nu}-g_{\mu\nu}\phi_{,\lambda}%
\phi_{,\kappa}\right]  +\frac{1}{4}\,(2\omega+1)\left(  g_{\lambda\nu}%
g_{\mu\kappa}-g_{\lambda\kappa}g_{\mu\nu}\right)  \phi_{,\alpha}\phi^{,\alpha}
\label{a9}%
\end{equation}
The equation (\ref{a8}) means that even in the absence of matter,\ except for
$\omega=-\frac{1}{2}$, the space-time is not necessarily Riemann flat as it
depends on the Weyl scalar field. Thus, unlike general relativity in (2+1)
dimensions, in WIST gravity the gravitational field does not necessarily
vanish outside the sources. Likewise, the curvature tensor $\mathcal{R}%
_{\ \mu\nu\kappa}^{\alpha}$ calculated with the Weyl connection $\Gamma
_{\beta\lambda}^{\alpha}$ does not vanish in the absence of matter as it is
given by%

\begin{equation}
\mathcal{R}_{\ \mu\nu\kappa}^{\alpha}=g^{\alpha\lambda}R_{\lambda\mu\nu\kappa
}-e^{-\phi}g_{\mu\lambda}\delta_{\lbrack\nu}^{[\alpha}Q_{\kappa]}^{\lambda]},
\label{a10}%
\end{equation}
with $Q_{\beta}^{\alpha}=4e^{\phi/2}\left(  e^{\phi/2}\right)  _{,\beta
;\lambda}g^{\alpha\lambda}-2\left(  e^{\phi/2}\right)  _{,\mu}\left(
e^{\phi/2}\right)  _{,\nu}g^{\mu\nu}\delta_{\,\beta}^{\alpha}$, and
$\delta_{\lbrack\nu}^{[\alpha}Q_{\kappa]}^{\lambda]}$ denotes
antisymmetrization with respect to both upper and lower indices. In the next
section, we shall investigate the Newtonian limit of Weyl gravity in the
weak-field regime.

\section{The Newtonian limit}

A metric theory of gravity is said to possess a Newtonian limit in the
non-relativistic weak-field regime if one can derive Newton's second law from
the geodesic equations as well as Poisson's equation from the gravitational
field equations. Let us now proceed to examine whether Weyl gravity fulfills
these requirements. The method we shall employ here to treat this problem is
standard and can be found in most textbooks on general relativity ( see, for
instance, ref. \cite{Bazin} ).

Since in Newtonian mechanics the space geometry is Euclidean, a weak
gravitational field in a geometric theory of gravity should manifest itself as
a metric phenomenon\ through a slight perturbation of the Minkowskian
space-time metric. Thus we consider a time-independent metric tensor of the
form
\begin{equation}
g_{\mu\nu}=\eta_{\mu\nu}+\epsilon h_{\mu\nu}, \label{quasi-Minkowskian}%
\end{equation}
\ where $n_{\mu\nu}$ denotes Minkowski metric tensor, $\epsilon$ is a small
parameter and the term $\epsilon h_{\mu\nu}$ represents a very small
time-independent perturbation due to the presence of some matter
configuration. Since we are working in the non-relativistic regime we shall
suppose that the velocity $V$ of a particle moving along a geodesic is much
less then $c$, so that the parameter $\beta=\frac{V}{c}$ will be regarded as
very small; hence in our calculations only first-order terms in $\epsilon$ and
$\beta$ will be kept. The same kind of approximation will be assumed to hold
with respect to the Weyl scalar field $\phi$, which will be supposed to be
static and very small, i.e., of the same order as $\epsilon$, and to emphasise
this fact we shall write $\phi=\epsilon\varphi$, where $\varphi$ is finite.

If we adopt the Galilean coordinates of special relativity we can write the
line element defined by (\ref{quasi-Minkowskian}) as
\[
ds^{2}=(dx^{0})^{2}-(dx^{1})^{2}-(dx^{2})^{2}-\epsilon h_{\mu\nu}dx^{\mu
}dx^{\nu},
\]
which leads, in our approximation, to
\begin{equation}
\left(  \frac{ds}{dt}\right)  ^{2}\cong c^{2}(1+\epsilon h_{00}) \label{dsdt}%
\end{equation}
We now apply the same approximation to the geodesic equations
\begin{equation}
\frac{d^{2}x^{\mu}}{ds^{2}}+\Gamma_{\;\alpha\beta}^{\mu}\frac{dx^{\alpha}}%
{ds}\frac{dx^{\beta}}{ds}=0, \label{geodesics}%
\end{equation}
recalling that the symbol $\Gamma_{\;\alpha\beta}^{\mu}$ represents the
components of the Weyl affine connection. From (\ref{Weylconnection}) it is
easy to see that, to first order in $\epsilon,$ we have
\begin{equation}
\Gamma_{\;\mu\nu}^{\alpha}=\frac{\epsilon}{2}n^{\alpha\lambda}[h_{\lambda
\mu,\nu}+h_{\lambda\nu,\mu}-h_{\mu\nu,\lambda}+n_{\mu\nu}\varphi_{,\lambda
}-n_{\lambda\mu}\varphi_{,\nu}-n_{\lambda\nu}\varphi_{,\mu}]
\label{Weylconnection2}%
\end{equation}
It is not difficult to see that, unless $\mu=\nu=0$, the product
$\Gamma_{\;\alpha\beta}^{\mu}\frac{dx^{\alpha}}{ds}\frac{dx^{\beta}}{ds}$ is
of order $\epsilon\beta$ or higher. In this way, the geodesic equations
(\ref{geodesics}) become, to first order in $\epsilon$ and $\beta$%
\[
\frac{d^{2}x^{\mu}}{ds^{2}}+\Gamma_{\;00}^{\mu}\left(  \frac{dx^{0}}%
{ds}\right)  ^{2}=0
\]
By taking into account (\ref{dsdt}) the above equation may be written as
\begin{equation}
\frac{d^{2}x^{\mu}}{dt^{2}}+c^{2}\Gamma_{\;00}^{\mu}=0
\label{equation-of-motion}%
\end{equation}
Clearly for $\mu=0$ the equation (\ref{equation-of-motion}) reduces to an
identity. On the other hand, if $\mu$ is a spatial index, a simple calculation
gives us $\Gamma_{\;00}^{i}=-\frac{\epsilon}{2}\eta^{ij}\frac{\partial
}{\partial x^{j}}(h_{00}-\varphi)$, hence the geodesic equation in this
approximation becomes, in three-dimensional vector notation
\[
\frac{d^{2}\overrightarrow{X}}{dt^{2}}=-\frac{\epsilon}{2}c^{2}\overrightarrow
{\nabla}(h_{00}-\varphi),
\]
which is simply Newton's equation of motion in a classical gravitational field
provided we identify the scalar gravitational potential as
\begin{equation}
U=\frac{\epsilon c^{2}}{2}(h_{00}-\varphi)\text{.} \label{Newtonian-potential}%
\end{equation}
It is interesting to note\ here the presence of the Weyl field $\varphi$ in
the equation above. It is the combination $h_{00}-\varphi$ that make up
the\ Newtonian potential$.$

Let us now turn our attention to the Newtonian limit of the field equations.
As we have seen previously, in $n$-dimensions the field equations of Weyl
gravity in the presence of matter are given by (\ref{Weyl-n}) and
(\ref{Weyl-n2}). It is now convenient to recast the equation (\ref{Weyl-n})
into the form
\begin{equation}
R_{\mu\nu}=-\kappa_{n}e^{(1-n/2)\phi}(T_{\mu\nu}-g_{\mu\nu}\frac{T}%
{n-2})-\frac{(n-1)(n-2)+4\omega}{4}\phi_{,\mu}\phi_{,\nu}
\label{Weyl-equations2}%
\end{equation}
In the weak-field approximation, i.e. when $g_{\mu\nu}=\eta_{\mu\nu}+\epsilon
h_{\mu\nu},$ it is easy to show that to first order in $\epsilon$, we have
$R_{00}=$ $-\frac{1}{2}\nabla^{2}\epsilon h_{00}$, where $\nabla^{2}$ denotes
the Laplacian operator in $n$-dimensional flat space-time. On the other hand,
because we are assuming a static regime $\phi_{,0}=0$, so the equation
(\ref{Weyl-equations2}) for $\mu=\nu=0$ now reads
\[
\nabla^{2}h_{00}=-\kappa_{n}\left(  T_{00}-\frac{g_{00}T}{n-2}\right)
e^{(1-n/2)\phi}%
\]
For a perfect fluid configuration, according to our minimal coupling
prescription, we have $T_{\mu\nu}=e^{-2\phi}[(\rho c^{2}+p)V_{\mu}V_{\nu
}-pg_{\mu\nu}]$, where $\rho$, $p$ and $V^{\mu}$ denotes, respectively, the
rest mass density, pressure and velocity field of the fluid. Since in this
approximation $\rho$ and $\phi=\epsilon\varphi$ are small quantities, we can
write $e^{-2\phi}\simeq$ $1-2\epsilon\varphi$, and thus, to first order in
$\epsilon,$ we have $T_{\mu\nu}\simeq\lbrack(\rho c^{2}+p)V_{\mu}V_{\nu
}-pg_{\mu\nu}]$. On the other hand, in a non-relativistic regime we can
neglect $p$ with respect to $\rho$, which \ then implies $T\simeq\rho c^{2}$.
Since in this approximation $\rho$ and $\phi=\epsilon\varphi$ are small
quantities, we have $Te^{(1-n/2)\phi}\simeq\rho c^{2}(1-\frac{\epsilon}%
{2}n\varphi)\simeq\rho c^{2}.$Thus we have
\begin{equation}
\frac{\epsilon}{2}\nabla^{2}h_{00}=\left(  \frac{n-3}{n-2}\right)  k_{n}\rho
c^{2} \label{A}%
\end{equation}
In the same approximation (\ref{Weyl-n}) becomes
\begin{equation}
\epsilon\nabla^{2}\varphi=\frac{-4\kappa_{n}\rho c^{2}}{(n-1)(n-2)+4\omega
}\text{ .} \label{B}%
\end{equation}
\ From (\ref{Newtonian-potential}), (\ref{A}) and (\ref{B}) we finally get the
n-dimensional Poisson's equation%

\begin{equation}
\nabla^{2}U=-K_{n}\rho\label{Poissonlike}%
\end{equation}
where $K_{n}=\kappa_{n}c^{4}\left(  \frac{n-3}{2(n-2)}+\frac{2}%
{(n-1)(n-2)+4\omega}\right)  $ plays the role of the gravitational constant in
$n$ dimensions. At this point we recall that in $n$-dimensional general
relativity the equation that corresponds to (\ref{Poissonlike}) is (see, for
instance, \cite{Giddings})%
\begin{equation}
\nabla^{2}U=\left(  \frac{n-3}{n-2}\right)  k_{n}\rho c^{4}. \label{PoissonGR}%
\end{equation}
\ For $n=3$ the right-hand side of the above equation vanishes, hence the
linearized Einstein theory fails to produce Newtonian gravity. However, due to
the presence of the scalar field, $K_{n}\neq$ $0$ . Therefore, WIST has a
Newtonian limit for any $n\geq3$.

\section{Geodesic deviation}

An aspect of the strange behaviour of three-dimensional general relativity,
first pointed out by Giddings et al (\cite{Giddings}), is the prediction that
world lines of dust particles do not deviate. This is equivalent to saying
that even if the space-time is allowed to have curvature these particles do
not feel the gravitational interaction. Let us now investigate the same
phenomenom in the light of Weyl integrable space-time theory.

Suppose that as the source of the gravitational field we have a pressureless
perfect fluid ("dust"). In this case, the energy-momentum tensor of the fluid
is given by
\begin{equation}
T^{\alpha\beta}=\rho u^{\alpha}u^{\beta},\label{energy-momentum}%
\end{equation}
where $u^{\alpha}=u^{\alpha}(x)$ denotes the components of the $4$-velocity
field of the fluid particles. Let $\eta^{\alpha}$ denote the deviation vector
of the congruence of geodesics determined by $u^{\alpha}$. The equation of
geodesic deviation is given by
\begin{equation}
\frac{D^{2}\eta^{\alpha}}{ds^{2}}=\mathcal{R}_{\;\mu\nu\kappa}^{\alpha}u^{\mu
}u^{\kappa}\eta^{\nu}\text{ },\label{Deviation}%
\end{equation}
where the operator $\frac{D}{ds}$ stands for the absolute derivative along the
geodesic congruence. Now, from (\ref{Weyl-equations2}) the identity
(\ref{Riemann}), the above equation may be written as
\begin{align}
\frac{D^{2}\eta^{\alpha}}{ds^{2}} &  =\left(  \frac{1+2\omega}{2}\right)
(\phi_{,\mu}\phi_{,\nu}u^{\mu}u^{\alpha}\eta^{\nu}-\phi_{,\mu}\phi_{,\kappa
}u^{\mu}u^{\kappa}\eta^{\alpha}-e^{\phi}\phi^{,\alpha}\phi_{,\nu}\eta^{\nu
}+\phi^{\alpha}\phi_{,\kappa}u_{\nu}u^{\kappa}\eta^{\nu}-\frac{1}{2}\phi
_{,\mu}\phi^{,\mu}u^{\alpha}u_{\nu}\eta^{\nu}\nonumber\\
&  +\frac{1}{2}e^{\phi}\phi_{,\mu}\phi^{\mu}\eta^{\alpha})-e^{-\phi}%
g_{\mu\lambda}\delta_{\lbrack\nu}^{[\alpha}Q_{\kappa]}^{\lambda]}u^{\mu
}u^{\kappa}\eta^{\nu}.
\end{align}
We thus see that in three-dimensional Weyl gravity the congruence of world
lines of particles of a pressureless fluid has a non-vanishing geodesic
deviation, so that a pair of \ freely falling particles will exhibit a
relative accelerated motion, revealing the presence of a gravitational field.
It is interesting to note that in this case the gravitational field manifests
itself only through the Weyl scalar field $\phi$.

\section{A Static and circularly symmetric solution}

In this section we consider the problem of determining the gravitational field
generated by a circularly symmetric matter distribution in a region outside
the source. By solving the field equations we obtain a vacuum static solution,
which, as we shall see, unlike three-dimensional general relativity, is not
flat in regions where matter is absent.

Let us start by writing the line element of a circularly symmetric space-time
in its most general form, which may be given by%

\begin{equation}
ds^{2}=e^{2N}dt^{2}-e^{2P}dr^{2}-r^{2}d\theta\text{,} \label{metric}%
\end{equation}
where $N(r)$ and $P(r)$ are functions of the radial coordinate only. It
is\ also assumed that the Weyl scalar field $\phi=\phi(r)$ also depends only
on $r.$Then, the field equations (\ref{a6}) and (\ref{fi}) becomes
\begin{equation}
N^{\prime\prime}+N^{\prime2}-N^{\prime}P^{\prime}+\frac{N^{\prime}}%
{r}=0\text{,} \label{eq1}%
\end{equation}

\begin{equation}
N^{\prime\prime}+N^{\prime2}-N^{\prime}P^{\prime}-\frac{P^{\prime}}{r}%
=\lambda\phi^{\prime2}\text{ ,} \label{eq2}%
\end{equation}%
\begin{equation}
N^{\prime}-P^{\prime}=0\text{ ,} \label{q3}%
\end{equation}%
\begin{equation}
\phi^{\prime\prime}+\phi^{\prime}(N^{\prime}-P^{\prime})+\frac{\phi^{\prime}%
}{r}=0\text{ ,} \label{q4}%
\end{equation}
\bigskip where prime denotes derivative with respect to $r$ and $\lambda
=-\frac{1}{2}(1+2\omega)$.

By substituting (\ref{q3}) into (\ref{q4}) we get%

\[
\phi^{\prime\prime}+\frac{\phi^{\prime}}{r}=0\text{ ,}%
\]
whose general solution is given by
\[
\phi=\phi_{0}+A\ln r\text{ ,}%
\]
where $\phi_{0}$ and $A$ are integration constants. (Since in Weyl geometry
the presence of an additive constant in the expression of the scalar field has
no geometrical meaning it is convenient to set $\phi_{0}=0$.) On the other
hand, also from (\ref{q3}), the equations (\ref{eq1}) and (\ref{eq2}) reduce,
respectively, to
\[
N^{\prime\prime}+\frac{N^{\prime}}{r}=0
\]
and
\[
N^{\prime\prime}-\frac{N^{\prime}}{r}=\lambda\phi^{\prime2}.
\]
It is easily seen that the above equations yield
\[
N=N_{0}+B\ln r\text{ ,}%
\]
with $B=-\frac{\lambda}{2}A^{2}$, while (\ref{q3}) gives
\[
P=P_{0}+B\ln r\text{ ,}%
\]
where $N_{0}$ and\ $P_{0}$ are\ integration constants.

By rescaling the time coordinate $t$ we can set $N_{0}=0$. On the other hand,
if we assume that there is no conical singularity in the space-time we can
also take $P_{0}=0.$ Finally, the line element (\ref{metric}) may be written as%

\begin{equation}
ds^{2}=r^{2B}dt^{2}-r^{2B}dr^{2}-r^{2}d\theta^{2}\text{ ,}
\label{final solution}%
\end{equation}
while the scalar field is given by
\begin{equation}
\phi=A\ln r\text{ .} \label{scalar field}%
\end{equation}
Let us now apply the weak field limit to find the constant $B$ in terms of the
mass $M$ of the matter distribution. To do this, we first note that when $B=0$
(\ref{final solution}) reduces to the metric of Minkowski space-time. From
(\ref{scalar field}) it seems natural to identify the parameter $\epsilon$ of
Section V with the constant $A.$ For small values of $B$ we can write
$r^{2B}=e^{\ln r^{2B}}\simeq1+2B\ln r$. At this point, let us recall that in
two spatial dimensions the Newtonian gravitational potential $U(r)$ of a
circularly symmetric mass distribution is given by $U(r)=G_{2}M\ln r$, where
$G_{2}$ denotes the gravitational constant in this dimensionality. Now, from
(\ref{Newtonian-potential}) and recalling that $B=-\frac{\lambda}{2}A^{2}$ is
a second-order term in $\epsilon=A$ we obtain $A=-\frac{2MG_{2}}{c^{2}}$ and
$B=-\frac{2\lambda M^{2}G_{2}^{2}}{c^{4}}=\frac{(1+2w)M^{2}G_{2}^{2}}{c^{4}}$.

With respect to the space-time corresponding to this solution a few comments
are in order. First, let us note that the basic invariants in this
dimensionality are $\ I_{1}=e^{\phi}\mathcal{R}$ and $I_{2}=e^{2\phi
}\mathcal{R}_{\mu\nu}\mathcal{R}^{\mu\nu}$. For the space-time
(\ref{final solution}) we have $I_{1}=\frac{4B-A^{2}}{2}r^{A-2B-2}$ and
$I_{2}=$ $F(A,B)$ $r^{2(A-2B-2)}$, where $F(A,B)$ is a function of the
constants $A$ and $B$. Thus, it is not difficult to verify that that this
solution corresponds to a space-time presenting a naked singularity at the
center of the matter distribution for values of $\omega>-\frac{1}{2}%
+\omega_{0}$, with $\omega_{0}=\frac{A-2}{A^{2}}$. In fact, this\ result is
not surprising since naked singularities are a common feature observed in
(3+1)-dimensional general relativity models with a massless scalar fields
\cite{Roberts}.

\section{Final remarks}

Investigation in lower dimensional gravity has arisen essentialy after the
realization that the two or three-dimensional versions of Einstein gravity are
rather peculiar and, in some sense, unsuitable for a theory of the
gravitational interaction (for instance, gravitational waves do not exist in
three-dimensional general relativity). However, the current motivation for
this kind of research is that lower dimensional models can provide useful
insights and ideas to construct a successful quantum theory of gravity
\cite{carlip}. Apart from general relativity, other theories of gravity have
been studied in this context. Of much interest is the study of how black holes
may form in gravity theories formulated in lower dimensions. Because general
relativistic space-times are locally flat outside matter sources there are no
black holes solutions, unless a negative cosmological constant is introduced,
as was pointed out by Ba\~{n}ados \textit{et al} some years ago \cite{Banados}%
. On the other hand, different models of dilaton gravity also lead to the
discovery of black hole solutions. In this work, we have approached the
subject from the standpoint of Weyl integrable geometry, in which a scalar
field plays the role of a geometrical field. We have found out that this
modification in the space-time geometry leads to new features that are not
present in three-dimensional Einstein gravity, such as the existence of a
Newtonian limit and the non-vanishing geodesic deviation of the trajectories
of \ freely falling particles. Finally, we have shown that another effect of
the geometrical scalar field is that the space-time generated by a static
circularly symmetric matter distribution corresponds, for certain values of
the parameter $\omega,$ to a curved space-time corresponding with a naked
singularity at the center of the distribution.

\bigskip

\bigskip

\section*{Acknowledgements}

\noindent The authors acknowledge Conacyt M\'{e}xico and CNPq-CLAF for
financial support.

\end{document}